# Real-Time Contingency Analysis with Corrective Transmission Switching —Part I: Methodology

Xingpeng Li, *Student Member, IEEE*, Pranavamoorthy Balasubramanian, *Student Member, IEEE*, Mostafa Sahraei-Ardakani, *Member, IEEE*, Mojdeh Abdi-Khorsand, *Student Member, IEEE*, Kory W. Hedman, *Member, IEEE*, and Robin Podmore, *Fellow, IEEE*

*Abstract*— Transmission switching (TS) has gained significant attention recently. However, barriers still remain and must be overcome before the technology can be adopted by the industry. The state of the art challenges include AC feasibility and performance, computational complexity, the ability to handle large-scale real power systems, and dynamic stability. This two-part paper investigates these challenges by developing an AC TS-based real-time contingency analysis (RTCA) tool that can handle large-scale systems within a reasonable time. The tool proposes multiple corrective switching actions, after detection of a contingency with potential violations. To reduce the computational complexity, three heuristic algorithms are proposed to generate a small set of candidates for switching. Parallel computing is implemented to further speed up the solution time. Furthermore, stability analysis is performed to check for dynamic stability of proposed TS solutions. Part I of the paper presents a comprehensive literature review and the methodology. The promising results, tested on the Tennessee Valley Authority (TVA) system and actual energy management system (EMS) snapshots from Pennsylvania New Jersey Maryland (PJM) and the Electric Reliability Council of Texas (ERCOT), are presented in Part II. It is concluded that RTCA with corrective TS significantly reduces potential post-contingency violations and is ripe for industry adoption.

## I. INTRODUCTION

**M**AINTAINING a reliable power system is of utmost importance. The North American Electric Reliability Corporation (NERC) requires power systems to withstand the loss of a single bulk electric element (*N-1*) [1]. While various classes of reserves are acquired, reliable operation is not always achieved. Real-time contingency analysis (RTCA) is frequently repeated for this purpose.

In the Midcontinent Independent System Operator (MISO) system, the RTCA package simulates more than 11,500 contingency scenarios every four minutes [2]. RTCA utilizes data from the state estimator and contingency analysis is performed by successively solving AC power flows. Thermal and voltage violations corresponding to different contingencies are, then, determined [3] by analyzing the power flow results.

Pennsylvania New Jersey Maryland (PJM) operators simulate a full AC contingency analysis to identify the contingencies that cause violations in the system [4]. Approximately 6,000 contingencies are assessed every minute at PJM [4]. Although there is a list of all contingencies in PJM's database, not all contingencies in that list are evaluated at all times [5].

The Electric Reliability Council of Texas (ERCOT) uses a two-phase procedure to perform breaker-to-breaker contingency analysis [6]. A heuristic screening procedure is performed in the first phase to identify the most severe contingencies based on the post-contingency violations. Previously, ERCOT had approximately 3938 contingencies, including 2958 single branch contingencies, 375 double branch contingencies, and 605 generator contingencies, modeled in its system [7]. The RTCA in ERCOT executes every five minutes [7].

If a contingency with post-contingency violations is detected, appropriate actions will be taken to ensure reliability. These actions include:
- Sending constraints to security-constrained economic dispatch to move away from a vulnerable state.
- Commitment of fast-start units to improve availability of local reserves.
- Transmission switching (TS) to enhance deliverability of reserves.

Corrective transmission switching (CTS) is shown to be a viable solution [8]-[10] for handling contingencies, which is also significantly cheaper than its alternatives. CTS is already being used in normal and post-contingency operation, though to a very limited extent, at PJM [11]. Despite the vast body of literature that has been dedicated to TS over the last decade, important challenges remain for more systematic adoption of the technology. The challenges include computational complexity, unknown or poor AC performance, concerns regarding the stability of switching actions, and limited insight on performance of the technology on actual large-scale power system data. This paper closes an important gap in the literature by addressing these challenges for the RTCA application. An open source AC CTS-based RTCA tool, which is fast and works with actual power system data, is developed. The tool is tested on the Tennessee Valley Authority (TVA) system and actual energy management system (EMS) snapshots from PJM and ERCOT. High performance computing (HPC) is em-

The research is funded by the Department of Energy (DOE) Advanced Projects Agency – Energy (ARPA-E) under the Green Electricity Network Integration (GENI) program.

X. Li, P. Balasubramanian, M. Sahraei-Ardakani, M. Abdi-Khorsand, and K. W. Hedman are with the School of Electrical, Computer, and Energy Engineering, Arizona State University, Tempe, AZ, 85287, USA (e-mail: {Xingpeng.li; pbalasu3; Mostafa; mabdikho}@asu.edu; kwh@myuw.net).

R. Podmore is the founder and president of IncSys, Bellevue, WA, 98007, USA, (e-mail: robin@incsys.com).

ployed to improve computational efficiency. The results, presented in Part II [12], are very promising and show that CTS can provide significant reliability benefits by drastically reducing the potential post-contingency violations. This will translate into significant savings due to substantially reduced need for expensive reliability-motivated generation redispatch and commitment. The tool is able to handle the PJM system in about five minutes with a standard desktop and parallel computing can be used to further reduce the solution time. Furthermore, stability analysis is performed on selected cases to test the switching solutions are dynamically stable.

The rest of this paper is organized as follows. Section II presents a literature review on TS and the challenges for its implementation. Section III explains the concept of corrective TS. Section IV presents the algorithm and methodology. Parallel computing details are presented in Section V. Stability analysis methodology is described in Section VI. Finally, Section VII concludes the paper.

## II. LITERATURE REVIEW

Currently, power system software does not utilize the flexibility of the transmission network and transmission elements are modeled as fixed assets. A single network topology is likely not optimal for different hours corresponding to different operating states of the system. Even though the flexibility in the transmission network is not modeled in optimal power flow and economic dispatch problems, it is well known that the system operators can and do change the network topology in practice [13]-[16].

Previous research has demonstrated that TS provides a variety of benefits including cost savings [17]-[18], active power loss reduction [19]-[20], thermal and voltage violation reduction [21]-[24], and enhancement of integration of renewable energy resources [25]. Furthermore, TS is shown to be beneficial in load shed recovery [26], enhancement of do-not-exceed limits [27]-[28], security and cost improvement in transmission and generation expansion planning [29], and potential cost saving in outage coordination [30].

It is illustrated in [31] that the optimal solution with TS will be at least as good as the solution obtained without TS. Co-optimization of unit commitment and TS is presented in [32]. Numerical studies show that the optimal network topology could be different for subsequent hours and that it is even possible to eliminate the need to commit additional generators as the deliverability of reserves is improved via TS. Tests conducted on the standard IEEE 118-bus test case demonstrate that 25% saving in system dispatch cost could be achieved by optimizing the transmission network topology [33]. It is a general concern that TS may compromise the reliability of the system. However, [34] shows that 15% of the overall cost can be reduced via optimizing the transmission topology, while still maintaining *N*-1 reliability.

TS is a power flow control technology that can improve the transfer capability and reduce the cost due to thermal and voltage limits. The total congestion costs in the PJM system in 2013 increased by $147.9 million, which amounts to a 28% increase compared to 2012 level of $529 million [35]. Therefore, there is a great opportunity for efficiency improvement through TS and other power flow control technologies, such as flexible AC transmission system (FACTS) devices [36]-[38].

A major advantage of TS is that it does not require installation of sophisticated hardware, such as expensive FACTS; TS can be performed with existing circuit breakers. Therefore, TS is a low cost power flow control technology that can significantly improve the efficiency of the power system. All the promising findings described above indicate that TS is an efficient and low-cost technology for building a smarter and more flexible electric grid.

Optimizing transmission line configurations is proposed as one of the advanced transmission technologies in the Energy Policy Act of 2005 [39]. The mathematical representation of optimal TS (OTS), with a DC set of assumptions, is a mixed integer linear program (MILP) with binary variables representing the status of the switchable transmission assets (line or transformer). Two different procedures, deterministic MILP and genetic algorithm, are proposed to find the best network topology for congestion management [40]. Results show that network reconfiguration could be used as an effective mechanism to relieve network congestion.

One of the main challenges facing industry adoption of TS is its computational complexity. MILPs are computationally expensive problems, specifically for the OTS application, considering the size of the power system. Thus, it is a challenge to get an exact solution for OTS with MILP-based within the limited available time. Therefore, many attempts to reduce the computational complexity of the problem have been pursued. In [41], a reduced MILP formulation is developed with power transfer distribution factors (PTDF) and line outage distribution factors [42]. Sensitivity analysis has been used as a way to overcome the computational complexity of the problem [43]-[46]. A greedy algorithm is proposed to quickly identify switching candidates using duals from a DC optimal power flow (DCOPF) [45] and an AC optimal power flow (ACOPF) [46]. However, [47] studies these algorithms on a large-scale Polish system and concluded that this heuristic may be inconsistent in performance. A cycle-based formulation for OTS is proposed in [48] to reduce the computational complexity by providing strong valid inequalities for the use of cutting-plane approach in a DC framework. Alternatively, high performance computing could be used to reduce the solution time for TS applications [49].

Due to the above mentioned reasons, as well as other concerns such as dynamic stability, implementation of TS has been very limited. Some system operators use TS as a corrective mechanism for improving voltage profiles and mitigating thermal overloads [14], [50]. TS is also being employed during planned outages, to make the transition smooth, and as a post-contingency corrective action [51]. California ISO (CAISO) is reported to perform TS on a seasonal basis and to relieve congestion in the system [14], [16], [52]. PJM has posted a list of potential transmission switching solutions that may reduce or eliminate violations for normal and post-contingency situations [11], [53]. However, these switching actions are not guaranteed to always provide benefits because they are identified offline; even when they do provide benefits, they may not be the best option [53]. The performance of a particular TS action is highly dependent on the operating state of the system.

This paper develops an RTCA package that effectively incorporates TS as a corrective mechanism. The major bottle-necks to the implementation of TS are addressed by developing a RTCA tool with corrective TS. Prior work on CTS focuses on approaches that do not confirm with the modelling requirements or are based on algorithms that do not scale well. The majority of literature on this topic is based on small-scale test systems with DC power flow models.

While the developed RTCA tool with CTS is straightforward from an algorithmic sense, it is for that reason why the approach is innovative and ready to make a large impact in industry; it is not only scalable but it is also highly effective and, thus, it bridges the gap between the existing technology and actual implementation. The contributions of this paper are the following:

1. The algorithms developed are extremely fast. In fact, they can handle a snapshot of PJM in about five minutes with a standard desktop. Parallel computing would further improve the solution time. Therefore, this paper effectively tackles the computational complexity of CTS.

2. Full AC power flow is used for implementation of the method. Therefore, there will be no ambiguity on the performance of the solution in an AC setting.

3. The tool is able to handle large-scale systems. The TVA system and actual snapshots from the EMS of PJM and ERCOT are used to test the tool.

4. Stability analysis is performed on a subset of cases using standard industry software such as PSS/E to test the dynamic stability of proposed solutions.

This paper studies CTS with the details explained above and addresses the state of the art challenges of CTS. Therefore, the conclusions presented in this paper are more comprehensive compared to earlier studies. Preliminary results and conclusions related to this work were published in [54]. This paper elaborates on [54] and presents the algorithm development. The paper also provides detailed information on high performance computing implementation as well as stability analysis. Part II of this paper presents detailed results obtained on the PJM, TVA, and ERCOT systems as well as stability analysis and high performance computing performance. Thus, this paper closes a significant gap in the literature by accomplishing the above-mentioned goals.

## III. CONCEPT OF CORRECTIVE SWITCHING

This section presents two examples to show how CTS can reduce or eliminate post-contingency violations. Fig. 1 shows an example that CTS fully eliminates all the voltage violations caused by a transmission contingency. This example is from the authors' prior work in [24]. The network shown in Fig. 1 is a 500KV level portion of the TVA system. This particular case corresponds to a lightly loaded period. In the pre-contingency state, the switching candidate produces reactive power, which travels through the contingency line. In the post-contingency state, too much reactive power must stay in the affected area since the contingency line is now not available to deliver the excessive reactive power out of this area, which leads to over voltage. In the post-switching state, the source element of producing that excessive reactive power, which is the CTS solution itself, is removed from the system; hence, the over voltage violations are all eliminated.

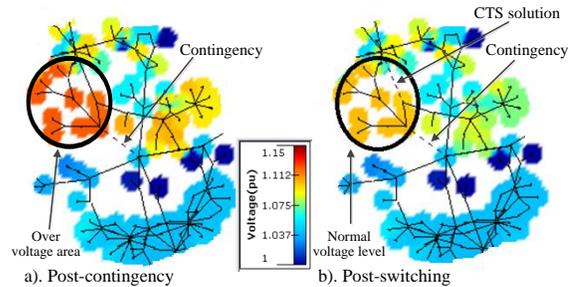

Fig. 1. An example of voltage violation fully eliminated by CTS; voltage contour plot.

Fig. 2 presents an example, which depicts how CTS can eliminate flow violations in the PJM system. In Fig. 2, bus 7 is the load pocket. A contingency on branch 1-5 overloads the line 3-4. Switching line 2-3 relieves the overload and the power flow is rerouted to bus 7 through the external circuit.

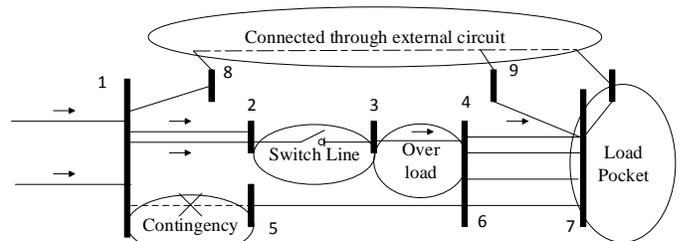

Fig. 2. Network diagram for CTS mechanism.

## IV. METHODOLOGY AND ALGORITHM

The procedure for contingency analysis with CTS is presented in Fig. 3. The normal operating state of the system, which consists of AC power flow information, is first fed into RTCA. Contingency analysis is then performed and the contingencies that cause violations beyond a specific threshold are identified. The tolerances used in this paper are 0.005 per unit for voltage violation and 5 MVA for flow violations. Both metrics are based on an aggregate level across the entire system. Only those contingencies with violations beyond the thresholds are sent to the CTS routine. Five switching candidates, which would eliminate or reduce the violations, are identified for each of those contingencies.

Stability analysis, using a standard industry tool such as PSS/E, is performed on selected cases to test for dynamic stability after performing the corrective switching action.

### A. Contingency Analysis

RTCA is a well-known and essential function in modern energy management systems. The RTCA package developed in this paper adopts the standard assumptions as given below:

1. For transmission element contingencies, all generators' active power outputs remain at the pre-contingency level except for the generators at the slack bus(es).

2. For generator contingencies, participation factor based on available capacity is used to redispatch generation [24].

3. The faulty element is isolated using circuit breakers.

This paper utilizes OpenPA [55], an open source decoupled AC power flow [56], as the power flow engine of the RTCA.

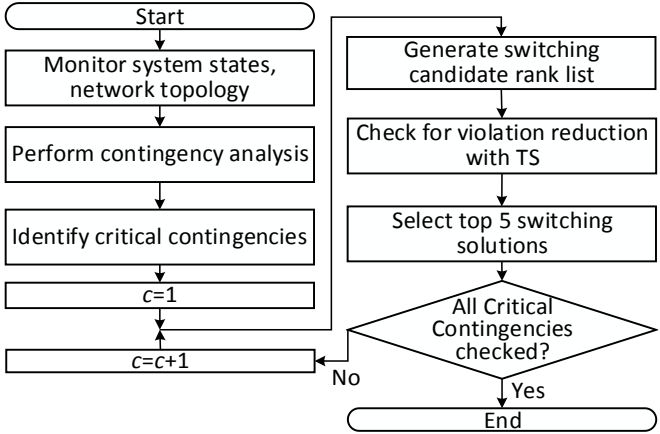

Fig. 3. Procedure for contingency analysis with CTS.

If the right corrective actions are not implemented, contingencies may cause post-contingency voltage limit violations and line overloading. Such conditions may lead to a cascading outage. A possible result of cascading failures is a system blackout, with disastrous socio-economic consequences. Therefore, managing post-contingency violations appropriately is essential for secure operation of the system.

Operators have several options to maintain reliability after detection of a contingency with potential violations. The operator can move the dispatch away from the vulnerable state or commit additional units. These are common means of maintaining reliability. Note that reliability motivated commitments and redispatch create a significant economic burden. While such economic burdens are justifiable in order to prevent catastrophic blackouts, there are cheaper solutions.

CTS is shown to be an effective alternative to many preventive approaches. Not only is CTS effective, it is also drastically cheaper. CTS may completely eliminate the potential post-contingency violations or significantly reduce them. Thus, there will be substantially less need for expensive reliability-motivated commitments and redispatch. Taking everything into account, corrective transmission switching provides reliability benefits, through which significant cost savings will be achieved.

Note that system operators do not model all potential *N*-1 events in RTCA. Various contingencies are not likely to cause violations based on the immediate system condition. The operator may also have a predetermined strategy to alleviate post-contingency violations, e.g., switchable shunts, adjustment of transformer taps, bus splitting, FACTS devices, or a predetermined special protection scheme (SPS). To provide a comprehensive study, this paper simulates all the potential *N*-1 events, excluding radial lines.

*B. Heuristic Approaches for Computational Tractability*

To reduce the computational complexity of the problem, three heuristics are proposed to generate a limited set of switching candidates. This fairly small subset of switchable elements includes quality solutions and also can be processed within a reasonable time, making the method suitable for real-time applications. The three heuristics, which are proposed in this paper to generate the candidate switching list, are listed below. Complete enumeration (CE) is only used to gauge the performance of the heuristics.

- Closest branches to contingency element (CBCE),
- Closest branches to violation elements (CBVE),
- Data mining (DM).

Based on the authors' prior experience, it is observed that most of the beneficial switching solutions lie within a close vicinity of the contingency element and/or the violations. Based on this observation, two heuristic approaches, CBCE and CBVE, are developed. CBCE searches for the 100 closest branches to the contingency element. CBVE heuristic searches for the 100 branches closest to the elements with violations.

For transmission contingencies, it is found that the network violations occur on the elements that are very close to the contingency element. Hence, the lists of transmission switching candidates generated by both the CBCE and CBVE would be very similar. For generator contingencies, since the generators are redispatched, the violations may not be that close to the contingency. In this case, it is very likely that the CBVE method provides better CTS solutions when compared with the CBCE heuristic.

The distance of one element to another element, used by both CBCE and CBVE in this paper, is defined as the number of branches in the shortest path connecting these two elements. Therefore, neither the *electrical distance* [57]-[58] nor the *real distance (miles)* is involved in this metric; the proximity of two elements is only determined by the topological characteristics of the network. Suppose the contingent element is a line; all other transmission assets that are directly connected to either of the two contingent line's buses, they are given a distance of zero. When the distance is listed as being zero for a line, the shortest path does not traverse across any other transmission asset to reach that specified line whose distance is zero. Lines that have a distance of one are connected to the far end bus of lines that have a distance of zero (for the corresponding shortest path). This process repeats to generate distances based purely on a topological structure and the shortest path.

For the data mining technique, it was first observed that many switching solutions come from a common subset of transmission assets. The system cases are split into two sets: training and test. Initially, complete enumeration of all the switchable branches is performed for each of the potential critical contingencies on the training set. The beneficial switching actions for each contingency in each scenario are identified and combined together. This combined list is a very small subset of all switchable elements. The combination of the beneficial switching actions for the training set is considered as the switching candidate list for the test set. The basic assumption behind this method is that, even if operational conditions change, previously determined beneficial switching solutions should be at least top candidates considered for the switching action. To be more specific, the training set can be considered historical information while the test set will consist of snapshots in real-time.

Different tolerances for identifying beneficial solutions with the DM method can result in different candidate list lengths. In Part II, three DM methods with different thresholds are studied. They are referred to as DM1, DM2, and DM3, respectively. There is no minimum threshold used in DM1 for identifying the beneficial switching solutions, which makes the list very long for this approach, since even the candidates produc-ing negligible improvements will be considered as potentially beneficial CTS solutions. Only the switching actions that provide a violation reduction of more than 5% comprise the candidates for CTS in DM2. DM3 has the smallest list length as it includes only those switching actions that provide a violation reduction of more than 10%.

Apart from the heuristic methods, a complete enumeration (CE) of all possible switching actions is also performed in order to estimate the best possible benefits that can be achieved with CTS. CE is obviously not a practical approach so it is merely used to confirm the optimal solution and to offer a basis for analysis of the quality of heuristic methods.

*C. Metrics*

Average violation reduction in percent is used to show the effectiveness of the method on an aggregate level:
$$P_{CTS} = \frac{1}{N_c}\sum_{c=1}^{N_c} \frac{(\Delta_{c0}-\Delta_{c1})}{\Delta_{c0}} * 100\% \tag{1}$$
where, $\Delta_{c0}$ denotes the total violations after contingency $c$; $\Delta_{c1}$ denotes the total violations after corrective switching with contingency $c$ still present; and $N_c$ is the total number of critical contingencies identified.

Although the post-contingency violations may be reduced on an aggregate level by implementing a specific CTS action, it is important to analyze the impact of the switching action on individual elements. It is possible that a specific switching action, while reducing the overall violations, creates additional violations that did not exist before implementation of the CTS action. CTS may also increase the violation on one particular element, while reducing the overall violations. Therefore, solutions are checked for Pareto improvements (PI); the CTS solution provides a Pareto improvement when at least one post-contingency violation reduces without causing any additional violations on any other element of the system.

Depth is defined as the location of the identified beneficial CTS solution in the candidate list. Depth is proposed only as a metric to evaluate the efficiency of each heuristic. The average depth can be calculated as follows:
$$D_{CTS} = \frac{1}{M_c}\sum_{c=1}^{M_c} L_{CTS,c} \tag{2}$$
where, $L_{CTS,c}$ denotes the location or the index of the identified beneficial CTS solution in the ranking list for contingency $c$; $M_c$ is the number of critical contingencies for which a beneficial CTS solution exists.

*D. N-1-1 Reliability*

Meeting the *N*-1-1 reliability requirement is essential to ensure a reliable system. Thus, one major CTS concern is regaining *N*-1 reliability after the first contingency and the related CTS action. NERC's *N*-1-1 reliability criterion states that the system has to become *N*-1 reliable again within 30 minutes following the first contingency. In the case of a contingency leading to network violations, a corrective action is first implemented to relieve the violations and bring the system back to acceptable operational conditions in a very short time. This paper proposes CTS as an effective corrective action at this initial step. Subsequently, remedial actions will be taken to regain *N*-1 reliability in post-switching situations. The remedial actions can include a mixture of generation redispatch, further CTS actions, and putting the switched CTS line back in service. This paper focuses on the immediate corrective action taken right after the occurrence of the contingency and *N*-1-1 reliability is outside the scope of this paper.

*E. Impacts of Switching Solutions on Circuit Breakers*

The proposed corrective TS scheme is intended to provide operators with additional corrective control actions that are very cheap and effective; the only associated cost of CTS is the impact on circuit breakers. CTS provides the operator with a corrective switching solution that would be implemented only if the contingency occurs. One concern regarding transmission switching is the impact on circuit breakers; ABB gave a presentation at PJM on circuit breaker health in relation to transmission switching, [59]. Since the probability of the contingency is low, the great benefit of this corrective control technology is that it would rarely need to be implemented. Additionally, there are many beneficial switching solutions, see [12], which makes the likelihood of multiple CTS actions on the same circuit breakers very low. Thus, the wear and tear on the circuit breaker due to CTS is minor.

*F. Multiple Switching Solutions*

Only a single switching action will be implemented as a corrective mechanism in this paper for each contingency. Although switching multiple lines simultaneously is theoretically possible and would provide more flexibility, we focus on single switching solutions based on the request of industry and we leave the investigation of multiple switching solutions to future work.

## V. HIGH PERFORMANCE COMPUTING

With the advancements in computer technology, high performance computing is increasingly gaining popularity. The hardware for implementation of HPC is easily available nowadays. Moreover, mature parallel programming models, such as message passing interface (MPI), make parallel programming attainable for real-world applications.

The nature of the CTS module within the developed RTCA is apt for parallel computing. Evaluation of each switching candidate is independent of other candidates and, thus, can be assigned to an independent processor. The problem is solved simultaneously with multiple threads by breaking it into independent sub-problems. Parallel computing enables such parallelizable problems to be solved significantly faster.

One metric to measure the effectiveness of parallel computing is the parallel efficiency as defined in (3),
$$\eta_n = T_1 / nT_n \tag{3}$$
where $n$ denotes the number of threads, $T_1$ denotes the computational time of the sequential program, and $T_n$ denotes the computational time of the parallel program with $n$ threads.

For RTCA with CTS, as presented in this paper, the parallel efficiency is close to one, suggesting significant room for computational efficiency gains via parallel computing.

The parallel computing tool used for the simulations performed in Part II of this paper is MPJ-Express [60], which is the message passing interface in JAVA. The hardware for parallel computing simulation used is "cab" cluster at Lawrence Livermore National Laboratory (LLNL).

## VI. STABILITY ANALYSIS

Power system stability is of utmost importance. It is not practical, and certainly not economically efficient, to design a system to be stable for all possible $N$-k contingency scenarios. Hence, the system is always designed such that it maintains stability only for a subset of disturbances around any given operating state.

The system stability, which is subject to a particular disturbance, depends on its current operating state and the nature of the disturbance. It is possible that the power system operating at a given set of conditions (operating states) is stable for a particular disturbance; however, the same disturbance may cause the system to collapse when it is operating at another particular stressed operating state. Hence, it is not proper to classify a disturbance as small or large depending on its magnitude alone.

Contingencies, as well as switching actions, are generally associated with large changes to the operating steady-state equilibrium of the system. Since the focus of this paper is contingency analysis with CTS, stability analysis plays an important role in this work. Moreover, there is an overarching concern that CTS may introduce more vulnerability to the system leading to system instability. These important issues are addressed through this paper.

Power system stability has been defined as the ability of the system for a given initial operating state to regain a state of operating equilibrium after occurrence of a physical disturbance with system variables remaining bounded [61]. Maintaining dynamic stability is an essential requirement for secure operation of the power system. Power system instability has been reported to cause several major blackouts in the past, which emphasizes the need to focus more on the power system stability studies [62].

Part II presents stability analysis results for corrective TS in PJM. The dynamic data for the PJM system contains information about the different machine models in the system. Time domain simulation is performed using PSS/E to analyze the effect of the proposed CTS actions on the system stability. The stability studies are conducted on selected hours with different loading profiles and different number of contingencies leading to network violations.

Time domain simulations are performed on all critical contingencies for the selected hours to check the stability of the switching solutions. It is very essential to check the stability of TS actions as unstable switching solutions would weaken the system rather than reducing the violations. Two different methodologies are followed to perform the time domain simulation for transmission contingencies and generator contingencies. In case of transmission contingencies, generation redispatch is not performed. The generators at the slack bus(es) are used to pick up the change in losses. However, generation redispatch based on the available capacity is implemented following a generator contingency.

For transmission contingencies, the base case power flow is run for the initial 2 seconds after which a transmission contingency is simulated. At $t=20$s, the CTS action is implemented and the simulation is terminated at $t=40$s.

The time domain simulation is run for a total of 60 seconds in case of generator contingencies. The base case is run for the initial 2 seconds without any disturbance to the system. The generation contingencies are simulated at $t=2$s and the generation redispatch associated with the particular contingency is implemented at $t=20$s. This is followed by the switching action, which is implemented at $t=40$s and the simulation is terminated at $t=60$s.

The rotor angle, frequency, and voltage stability are checked for the selected switching actions. The relative rotor angles of all machines are monitored throughout the duration of the simulation to ensure that no single machine or group of machines swing away from the rest of the system and lose synchronism. If there is a relative rotor angle separation of any machine from the rest of the system such that it loses synchronism, the CTS action is categorized as unstable. The frequency of all the buses in the area of disturbance is monitored and it is checked that the frequency stays within the limits of $59.5Hz<f<60.5Hz$ [63]. For any bus in the system, if the frequency deviates beyond the specified threshold, the switching action is considered to be unstable. Similarly, a voltage threshold of $0.9<V<1.1$ [63] is used to ensure that the switching action does not cause voltage instability.

Note that the objective of performing stability studies in this paper is to check if the switching solution is stable, assuming the system remains stable after the contingency. Hence, the emphasis of this study is on the stability of CTS action, not the dynamics of the contingency itself. The results presented in Part II show that the majority of the proposed switching candidates that are tested are stable, suggesting that TS is a viable corrective mechanism. This is consistent with how PJM views the stability issues in its footprint [64]. PJM concludes that stability has yet to become a significant system limitation [64].

## VII. CONCLUSIONS

Transmission switching is a low cost power flow control technology that would reduce the operational cost, improve system reliability, and enhance integration of intermittent renewable resources. Despite all these benefits, the industry adoption of the technology has been fairly limited due to the following barriers: computational complexity for large-scale systems, ambiguous AC performance, and stability concerns.

This paper comprehensively addresses the state of the art challenges of transmission switching. An open source multi-threaded AC-based RTCA package, which incorporates CTS as a corrective mechanism, is developed. When RTCA identifies a contingency with potential network violations, a separate routine finds effective CTS actions to relieve the violations. Thus, the need for reliability-motivated commitment and redispatch will be drastically reduced. This will translate into reliability benefits and also substantial cost savings.

The RTCA developed in this paper proposes multiple switching actions for each contingency. For each potential switching solution, the violation reduction and a flag indicating a Pareto improvement is communicated to the operator. The operator has the choice to implement any of the solutions based on the associated violation reductions, Pareto performance, or stability concerns.

Part II presents the performance of the developed RTCA with actual data from PJM, ERCOT, and TVA. The results demonstrate the effectiveness of CTS based on its ability to reduce post-contingency violations in an AC setting.

ACKNOWLEDGEMENT

The authors would like to thank Dr. Deepak Rajan for his support with HPC implementation and Lawrence Livermore National Lab. Discussions with Dr. Vijay Vittal on stability issues, as well as feedback from Chris Mosier for the RTCA tool development, are greatly appreciated. The authors would also like to express thanks for valuable feedback from PJM, ERCOT, TVA, ISONE, MISO, NYISO, CAISO, Alstom Grid, ABB, Siemens, PECO, FP&L, Southern Co., National Grid (UK), RTE (France), and FERC.

OPEN SOURCE SOFTWARE ACCESS

The proposed technology was built around IncSys' and PowerData's open source decoupled power flow; interested parties can download the software [55]. For the proposed real-time contingency analysis with corrective transmission switching, which is implemented with MPI based HPC, interested parties can email Dr. Kory W. Hedman (kwh@myuw.net).

REFERENCES

[1] NERC, "Standard TPL-002-0b – system performance following loss of a single bulk electric system element," [Online]. Available: http://www.nerc.com/files/TPL-002-0b.pdf.
[2] MISO, "MISO reliability assurance," [Online]. Available: https://www.misoenergy.org/WhatWeDo/Pages/Reliability.aspx.
[3] MISO, "MISO's existing methods for managing voltage and plans to improve voltage profiles," [Online]. Available: http://www.ferc.gov/CalendarFiles/20120503131554-MISO.pdf
[4] J. Baranowski and D. J. French, "Operational use of contingency analysis at PJM," *IEEE PES General Meeting*, San Diego, CA, Jul. 2012.
[5] PJM, "LMP model information," [Online]. Available: http://www.pjm.com/markets-and-operations/energy/lmp-model-info.aspx.
[6] C. Thompson, K. McIntyre, S. Nuthalapati, and A. Garcia, "Real-time contingency analysis methods to mitigate congestion in the ERCOT region," *IEEE PES General Meeting*, Calgary, AB, Jul. 2009.
[7] F. Garcia, N. D. R. Sarma, V. Kanduri, and G. Nissankala, "ERCOT control center experience in using real-time contingency analysis in the new nodal market," *IEEE PES General Meeting*, San Diego, CA, Jul. 2012.
[8] A. S. Korad and K. W. Hedman, "Robust corrective topology control for system reliability," *IEEE Trans. Power Syst.*, vol. 28, no. 4, pp. 4042-4051, Nov. 2013.
[9] P. Balasubramanian and K. W. Hedman, "Real-time corrective switching in response to simultaneous contingencies," *J. Energy Eng.*, vol. vol. 14, no. 1, special issue: smart grid and emerging technology integration, Feb. 2014.
[10] M. Abdi-Khorsand and K. W. Hedman, "Day-ahead corrective transmission topology control," *IEEE PES General Meeting*, Washington D.C., USA, Jul. 2014.
[11] PJM, "Switching solutions," [Online]. Available: http://www.pjm.com/markets-and-operations/etools/oasis/system-information/switching-solutions.aspx.
[12] X. Li, M. Sahraei-Ardakani, P. Balasubramanian, M. Abdi-Khorsand, K. W. Hedman, and R. Podmore, "Real-time contingency analysis with corrective transmission switching- Part II: results and discussion," *IEEE Trans. Power Syst.*, under review.
[13] K. W. Hedman, R. P. O'Neill, E. B. Fisher, and S. S. Oren, "Optimal transmission switching-sensitivity analysis and extensions," *IEEE Trans. Power Syst.*, vol. 23, no. 3, pp. 1469-1479, Aug. 2008.
[14] California ISO, "Minimum effective threshold report," Mar. 2010, [Online]. Available: http://www.caiso.com/274c/274ce77df630.pdf.
[15] PJM, "Manual 3: transmission operations - section 5: index and operating procedures for PJM RTO operation," Jun. 2010, [Online]. Available: http://www.pjm.com/~/media/training/nerc-certifications/m03v37-transmission-operations.ashx.
[16] K. W. Hedman, S. S. Oren, and R. P. O'Neill, "A review of transmission switching and network topology optimization," *IEEE PES General Meeting*, Detroit, MI, Jul. 2011.
[17] A. Khodaei and M. Shahidehpour, "Transmission switching in security-constrained unit commitment," *IEEE Trans. Power Syst.*, vol. 25, no. 4, pp. 1937-1945, Nov. 2010.
[18] J. Han and A. Papavasiliou, "The impacts of transmission topology control on the European electricity network," *IEEE Trans. Power Syst.*, vol. 31, no. 1, pp. 496-507, Jan. 2016.
[19] R. Bacher and H. Glavitsch, "Loss reduction by network switching," *IEEE Trans. Power Syst.,* vol 3, no. 2, pp. 447-454, May 1988.
[20] G. Schnyder, and H. Glavitsch, "Security enhancement using an optimal switching power flow," *IEEE Trans. Power Syst.*, vol. 5, no. 2, pp. 674-681, May 1990.
[21] W. Shao and V. Vittal. "Corrective switching algorithm for relieving overloads and voltage violations," *IEEE Trans. Power Syst.*, vol. 20, no. 4, pp. 1877-1885, Nov. 2005.
[22] A. A. Mazi, B. F. Wollenberg, and M. H. Hesse, "Corrective control of power system flows by line and bus-bar switching," *IEEE Trans. Power Syst.*, vol. 1, no. 3, pp. 258-265, Aug. 1986.
[23] A. G. Bakirtzis and A. P. S. Meliopoulos, "Incorporation of switching operations in power system corrective control computations," *IEEE Trans. Power Syst.*, vol. 2, no. 3, pp. 669-676, Aug. 1987.
[24] X. Li, P. Balasubramanian, M. Abdi-Khorsand, A. S. Korad, and K. W. Hedman, "Effect of topology control on system reliability: TVA test case," *CIGRE US National Committee Grid of the Future Symposium*, Oct. 2014.
[25] F. Qiu and J. Wang, "Chance-constrained transmission switching with guaranteed wind power utilization," *IEEE Trans. Power Syst.*, vol. 30, no. 3, pp. 1270-1278, May 2015.
[26] A. R. Escobedo, E. Moreno-Centeno, and K. W. Hedman, "Topology control for load shed recovery," *IEEE Trans. Power Syst.*, vol. 2, no. 2, pp. 908-916, Mar. 2014.
[27] A. S. Korad and K. W. Hedman, "Enhancement of do-not-exceed limits with robust corrective topology control," *IEEE Trans. Power Syst.*, early access, Jul. 2015.
[28] A. S. Korad and K. W. Hedman. "Zonal do-not-exceed limits with robust corrective topology control," *ELECTR POW SYST RES*, vol. 129, pp. 235-242, Dec. 2015.
[29] A. Khodaei, M. Shahidehpour, and S. Kamalinia, "Transmission switching in expansion planning," *IEEE Trans. Power Syst.*, vol. 25, no. 3, pp. 1722-1733, Aug. 2010.
[30] X. Li and K. W. Hedman. "Fast heuristics for transmission outage coordination," *19th Power Systems Computation Conference (PSCC)*, accepted for publication, Genoa, Italy, Jun. 2016.
[31] A. S. Korad, P. Balasubramanian, and K. W. Hedman, "Robust corrective topology control," *Handbook of Clean Energy Systems*, pp. 1-17, Jul. 2015.
[32] K. W. Hedman, M. C. Ferris, R. P. O'Neill, E. B Fisher, and S. S. Oren, "Co-optimization of generation unit commitment and transmission switching with N-1 reliability," *IEEE Trans. Power Syst.*, vol. 25, no. 2, pp. 1052-1063, May 2010.
[33] E. B. Fisher, R. P. O'Neill, and M. C. Ferris, "Optimal transmission switching," *IEEE Trans. Power Syst.*, vol. 23, no. 3, pp. 1346-1355, Aug. 2008.
[34] K. W. Hedman, R. P. O'Neill, E. B. Fisher, and S. S. Oren, "Optimal transmission switching with contingency analysis," *IEEE Trans. Power Syst.*, vol. 24, no. 3, pp. 1577-1586, Aug. 2009.
[35] PJM, "PJM State of the Market," 2013, [Online]. Available: http://www.monitoringanalytics.com/reports/PJM_State_of_the_Market/2013.shtml.
[36] M. Sahraei-Ardakani and K. W. Hedman, "A fast linear programming approach for enhanced utilization of FACTS devices," *IEEE Trans. Power Syst.,* early access, Jul. 2015.
[37] M. Sahraei-Ardakani and K. W. Hedman, "Day-ahead corrective adjustment of FACTS reactance: a linear programming approach," *IEEE Trans. Power Syst.,* early access, Sep. 2015.
[38] M. Sahraei-Ardakani and S. Blumsack, "Transfer capability improvement through market-based operation of series FACTS devices," *IEEE Trans. Power Syst.,* under review, 2015.
[39] The USA Energy Policy Act of 2005 (Section 1223), PUBLIC LAW 109-58—AUG. 8, 2005, [Online]. Available: http://energy.gov/sites/prod/files/2013/10/f3/epact_2005.pdf
[40] G. Granelli, M. Montagna, F. Zanellini, P. Bresesyi, R. Vailati, and M. Innorta, "Optimal network reconfiguration for congestion management by deterministic and genetic algorithms," *Electric Power Systems Research*, vol. 76, no. 6, pp. 549-556, Apr. 2006.

[41] P. A. Ruiz, A. Rudkevich, M. C. Caramanis, E. Goldis, E. Ntakou, and C. R. Philbrick, "Reduced MIP formulation for transmission topology control," *50th Allerton Conference,* pp. 1073-1079. Oct. 2012.

[42] T. Guler, G. Gross, and M. Liu, "Generalized line outage distribution factors," *IEEE Trans. Power Syst.*, vol. 22, no. 2, pp. 879-881, May 2007.

[43] P. A. Ruiz, J. M. Foster, A. Rudkevich, and M. C. Caramanis. "On fast transmission topology control heuristics," *IEEE PES General Meeting*, Detroit, MI, Jul. 2011.

[44] P. A. Ruiz, J. M. Foster, A. Rudkevich, and M. C. Caramanis, "Tractable transmission topology control using sensitivity analysis," *IEEE Trans. Power Syst.*, vol. 27, no. 3, pp. 1550-1559, Aug. 2012.

[45] J. D. Fuller, R. Ramasra, and A. Cha. "Fast heuristics for transmission-line switching," *IEEE Trans. Power Syst.*, vol. 27, no. 3, pp. 1377-1386, Aug. 2012.

[46] M. Soroush and J. D. Fuller, "Accuracies of optimal transmission switching heuristics based on DCOPF and ACOPF," *IEEE Trans. Power Syst.*, vol. 29, no. 2, pp. 924-932, Mar. 2014.

[47] M. Sahraei-Ardakani, A. Korad, K. W. Hedman, P. Lipka, and S. Oren, "Performance of AC and DC based transmission switching heuristics on a large-scale Polish system," *IEEE PES General Meeting*, Washington D.C., Jul. 2014.

[48] B. Kocuk, H. Jeon, S. S. Dey, J. Linderoth, J. Luedtke, and A. Sun, "A cycle-based formulation and valid inequalities for DC power transmission problems with switching," *arXiv preprint arXiv*:1412.6245, v3, Oct. 2015.

[49] A. Papavasiliou, S. S. Oren, Z. Yang, P. Balasubramanian, and K. W. Hedman, "An application of high performance computing to transmission switching," *IREP Symposium- Bulk Power System Dynamics and Control-IX*, Rethymnon, Greece, Aug. 2013.

[50] ISO-NE, "ISO New England operating procedure No. 19: transmission operations," [Online]. Available: http://www.iso-ne.com/rules_proceds/operating/isone/op19/op19_rto_final.pdf.

[51] MISO, "Optimal transmission switching and RTO needs," [Online]. Available at: http://www.pjm.com/~/media/committees-groups/stakeholder-meetings/transmission-topology-control/20131119-item-02a-nivad-optimal-transmission-switching-andrto-needs-discussion-points.ashx.

[52] California ISO, "Transmission constraint relaxation parameter revision," ISO draft final proposal, Nov. 2012, [Online]. Available: http://www.caiso.com/Documents/StrawProposal-TransmissionConstraintRelaxationParameterChange.pdf

[53] P. Balasubramanian, M. Sahraei-Ardakani, X. Li, and K. W. Hedman, "Towards smart corrective switching: analysis and advancement of PJM's switching solutions," *IET Generation, Transmission, and Distribution*, accepted for publication.

[54] M. Sahraei-Ardakani, X. Li, P. Balasubramanian, K. W. Hedman, and M. Abdi-Khorsand, "Real-time contingency analysis with transmission switching on real power system data," *IEEE Power Engineering Letters*, pp. 1-2, accepted for publication, Aug. 2015.

[55] IncSys, "Open source power applications and utilities," [Online]. Available: https://github.com/powerdata/com.powerdata.openpa.

[56] B. Stott and O. Alsac, "Fast decoupled load flow," *IEEE Trans. Power App. Syst.*, vol. PAS-93, no. 3, pp. 859–869, May 1974.

[57] E. Cotilla-Sanchez, P. Hines, C. Barrows, and S. Blumsack, "Comparing the topological and electrical structure of the North American electric power infrastructure," *IEEE Syst. J.*, vol. 6, no. 4, pp. 616-626, Dec. 2012.

[58] E. Cotilla-Sanchez, P. Hines, C. Barrows, S. Blumsack, and M. Patel, "Multi-attribute partitioning of power networks based on electrical distance," *IEEE Trans. Power Syst.*, vol. 28, no. 4, pp. 4979-4987, Nov. 2013.

[59] ABB, "PJM: transmission topology control – circuit breaker reliability and maintenance," [Online]. Available: http://www.pjm.com/~/media/committees-groups/stakeholder-meetings/transmission-topology-control/20131119-item-05b-lane-circuit-breaker-reliability-and-maintenance.ashx.

[60] A. Shafi, B. Carpenter, and M. Baker, "Nested parallelism for multi-core HPC systems using Java," *J. Parallel Distrib. Comput.*, vol. 69, no. 6, pp. 532-545, Jun. 2009.

[61] P. Kundur, J. Paserba, V. Ajjarapu, G. Andersson, A. Bose, C. Canizares, N. Hatziargyriou, D. Hill, A. Stankovic, C. Taylor, T. Van Cutsem, and V. Vittal, "Definition and classification of power system stability IEEE/CIGRE joint task force on stability terms and definitions," *IEEE Trans. Power Syst.*, vol. 19, no.2, pp. 1387-1401, Aug. 2004.

[62] G. S. Vassell, "Northeast blackout of 1965," *IEEE Power Engineering Review*, vol. 11, no. 1, pp. 4-8, Jan. 1991.

[63] NERC, "Standard PRC-024-1-generator frequency and voltage protective relay settings," [Online]. Available: http://www.nerc.com/_layouts/PrintStandard.aspx?standardnumber=PRC-024-1&title=Generator%20Frequency%20and%20Voltage%20Protective%20Relay%20Settings&jurisdiction=United%20States.

[64] PJM, "Fundamentals of transmission operations - system stability," [Online]. Available: http://www.pjm.com/~/media/training/new-pjm-cert-exams/foto-lesson6-stability.ashx.

**Xingpeng Li** (S'13) received his B.S. and M.S. degrees in electrical engineering in 2010 and 2013 from Shandong University and Zhejiang University, China, respectively. He is currently pursuing the Ph.D. degree in electrical engineering and the M.S. degree in industrial engineering at Arizona State University, Tempe, AZ, USA. His research interests include power system operations, planning, and optimization, microgrids, energy markets, and smart grid. He previously worked with ISO New England, Holyoke, MA, USA.

**Pranavamoorthy Balasubramanian** (S'08) received the B.E. degree in electrical and electronics engineering and M.E. degree in instrumentation engineering in 2009 and 2011 respectively from Anna University, Chennai, TN, India. He is currently working towards the Ph.D. degree in electrical engineering at Arizona State University. His research interests include power system operations and control, energy markets, smart grids, micro-grids, electrical machines, renewable energy sources, and process control.

**Mostafa Sahraei-Ardakani** (M'06) received the PhD degree in energy engineering from Pennsylvania State University, University Park, PA in 2013. He also holds the B.S. and M.S. degrees in electrical engineering from University of Tehran, Iran. Currently, he is a post-doctoral scholar in the School Electrical, Computer, and Energy Engineering at Arizona State University. His research interests include energy economics and policy, electricity markets, power system optimization, transmission network modeling, and smart grids.

**Mojdeh Abdi-Khorsand** (S'10) received the B.S. degree in electrical engineering from Mazandaran University, Babol, Iran, in 2007 and the M.S. degree in electrical engineering from the Iran University of Science and Technology, Tehran, Iran, in 2010. She is currently pursuing a Ph.D. degree and working as a research associate in the School of Electrical, Computer, and Energy Engineering at Arizona State University. Her research interests include power system operations, electricity markets, transmission switching, transient stability study and power system protection.

**Kory W. Hedman** (S' 05, M' 10) received the B.S. degree in electrical engineering and the B.S. degree in economics from the University of Washington, Seattle, in 2004 and the M.S. degree in economics and the M.S. degree in electrical engineering from Iowa State University, Ames, in 2006 and 2007, respectively. He received the M.S. and Ph.D. degrees in industrial engineering and operations research from the University of California, Berkeley in 2008 and 2010 respectively.

Currently, he is an assistant professor in the School of Electrical, Computer, and Energy Engineering at Arizona State University. He previously worked for the California ISO (CAISO), Folsom, CA, on transmission planning and he has worked with the Federal Energy Regulatory Commission (FERC), Washington, DC, on transmission switching. His research interests include power systems operations and planning, electricity markets, power systems economics, renewable energy, and operations research.

**Robin Podmore (M'73, F'96)** received the Bachelors and Ph.D. degrees in Electrical Engineering from University of Canterbury, N.Z in 1968 and 1973. In 1973 he worked as a Post-doctoral fellow at University of Saskatchewan, Saskatoon Canada. From 1974 to 1978 he managed the Power Systems Research group at Systems Control, Palo Alto, CA. From 1979 to July 1990 he was director and Vice President of Business Development with ESCA Corporation (now Alstom Grid), Bellevue WA. In July 1990 he founded and is President of Incremental Systems Corporation. He has been an industry leader and champion for open energy management systems, the Common Information Model, affordable and usable Operator Training Simulators and now affordable energy solutions for developing communities. He is a licensed Professional Engineer in the state of California. He is IEEE PES Vice President of New Initiatives and Outreach. Robin Podmore is a member of the USA National Academy of Engineering.